\documentclass[aps,prl,twocolumn,preprintnumbers,superscriptaddress,
              showpacs,nofootinbib]{revtex4-1}
\newcommand{\PRE}[1]{}       

\usepackage{bm} \usepackage{epsfig}

\usepackage{hyperref}

\usepackage{bm}

\def\beq{\begin{eqnarray}}
\def\eeq{\end{eqnarray}}
\def\bea{\begin{eqnarray}}
\def\eea{\end{eqnarray}}

\newcommand{\ifb}{\text{fb}^{-1}}

\newcommand{\gev}{\text{GeV}}
\newcommand{\tev}{\text{TeV}}

\newcommand{\eg}{{\em e.g.}}
\newcommand{\ie}{{\em i.e.}}

\newcommand{\eqref}[1]{Eq.~(\ref{#1})}

\newcommand{\figref}[1]{Fig.~\ref{fig:#1}}

\newcommand{\ssection}[1]{{\em #1\ }}

\newcommand{\drbar}{\ensuremath{\overline{\mbox{\sc dr}}}}
\newcommand{\msbar}{\ensuremath{\overline{\mbox{\sc ms}}}}
\newcommand{\hthreem}{{\sc H3m}}
\newcommand{\feynhiggs}{{\sc FeynHiggs}}
\newcommand{\softsusy}{{\sc SOFTSUSY}}
\newcommand{\suspect}{{\sc SuSpect}}
\newcommand{\spheno}{{\sc SPheno}}

\begin{document}

\preprint{UCI-TR-2013-08, HU-EP-13/28, SFB/CPP-13-37, CALT 68-2939}

\title{ \PRE{\vspace*{1.5in}} Three-Loop Corrections to the Higgs Boson
  Mass\\ and Implications for Supersymmetry at the LHC
  \PRE{\vspace*{0.3in}} }

\author{Jonathan L.~Feng}
\affiliation{Department of Physics and Astronomy, University of
  California, Irvine, CA 92697, USA}

\author{Philipp Kant}
\affiliation{Humboldt-Universit\"at zu Berlin, 12489 Berlin, Germany} 

\author{Stefano Profumo}
\affiliation{Department of Physics, University of California, 1156
  High Street, Santa Cruz, CA 95064, USA}
\affiliation{Santa Cruz Institute for Particle Physics, Santa Cruz, CA
  95064, USA}

\author{David Sanford
\PRE{\vspace*{.3in}}}
\affiliation{California Institute of Technology, Pasadena, CA 91125,
  USA
\PRE{\vspace*{.5in}}}

\begin{abstract}
\PRE{\vspace*{.3in}} In supersymmetric models with minimal particle
content and without left-right squark mixing, the conventional wisdom
is that the 125.6 GeV Higgs boson mass implies top squark masses of
$\mathcal{O}(10)~\tev$, far beyond the reach of colliders.  This
conclusion is subject to significant theoretical uncertainties,
however, and we provide evidence that it may be far too pessimistic.
We evaluate the Higgs boson mass, including the dominant three-loop
terms at $\mathcal{O}(\alpha_t \alpha_s^2)$, in currently viable
models.  For multi-TeV stops, the three-loop corrections can increase
the Higgs boson mass by as much as 3 GeV and lower the required stop
mass to 3 to 4 TeV, greatly improving prospects for supersymmetry
discovery at the upcoming run of the LHC and its high-luminosity
upgrade.
\end{abstract}


\maketitle

\ssection{Introduction.} The Higgs boson, recently discovered at the
LHC by the ATLAS and CMS Collaborations~\cite{:2012gk,:2012gu}, is now
the subject of impressive precision studies. In particular, combining
the results of all channels, the
currently available data, consisting of $25~\ifb$ collected at
$\sqrt{s} = 7$ and 8 TeV, constrain the Higgs boson mass to be
\begin{eqnarray}
\text{ATLAS (combined)} : && \! \! \! 125.5 \pm 0.2{~}^{+0.5}_{-0.6}~\gev 
~\mbox{\cite{ATLAS-CONF-2013-014}} \\
\text{CMS (combined)}:  && \! \! \!   125.7 \pm 0.3 \pm 0.3~\gev
~\mbox{\cite{CMS-PAS-HIG-13-005}} \ ,
\end{eqnarray}
where the first uncertainties are statistical and the second
systematic. Because the Higgs boson has been seen in purely leptonic
and photonic channels without missing $E_T$, its mass is already known
with a fractional uncertainty smaller than any of the quarks,
providing a potentially stringent bound on ideas for new physics.

The Higgs mass measurement is especially important for supersymmetry.
In supersymmetry, the Higgs quartic coupling is determined, at tree
level, by the gauge couplings, removing this {\em a priori} free
Standard Model parameter. The Higgs mass $m_h$ also receives large
radiative corrections, which are functions of superpartner masses. As
a result, $m_h$ provides useful guidance as to the mass scale of the
superpartners, with implications for direct discovery prospects for
supersymmetry at colliders.  Unfortunately, this potential is
currently clouded by theoretical uncertainties in the Higgs boson mass
calculation, which are arguably much larger than the experimental
uncertainties.  In this study, we extend previous work by including the
dominant 3-loop contributions to $m_h$ derived in
Refs.~\cite{Kant:2010tf,Harlander:2008ju}, and we explore implications
for supersymmetry discovery prospects at the LHC.

\ssection{The Higgs Mass at 3-Loops.}  In supersymmetric models with
minimal field content, the tree-level Higgs boson mass cannot exceed
$m_Z \simeq 91~\gev$.  The 1-loop contributions were explored long
ago~\cite{Okada:1990vk,Haber:1990aw,Ellis:1990nz}, and many studies
now incorporate 2-loop contributions, available with public codes such
as \feynhiggs~\cite{Heinemeyer:1998yj,Heinemeyer:1998np,%
  Degrassi:2002fi,Frank:2006yh}, \softsusy~\cite{Allanach:2001kg},
\suspect~\cite{Djouadi:2002ze}, and
\spheno~\cite{Porod:2003um,Porod:2011nf}.

The radiative corrections to the Higgs boson mass are most sensitive
to the top squark sector.  At tree-level, the top squark mass matrix
is
\begin{equation}
\left( \tilde{t}^*_L , \tilde{t}^*_R \right)
\! \! \left( \begin{array}{cc}
\! \! m_{\tilde{t}_L}^2 \! \! + m_t^2 + \Delta_L & m_t X_t \! \! \! \\ 
m_t X_t & \! \! \! m_{\tilde{t}_R}^2 \! \! + m_t^2 + \Delta_R \! \!
\end{array} \right) 
\! \! \left( \! \begin{array}{c}
\tilde{t}_L \\ \tilde{t}_R \end{array} \! \right) \! ,
\end{equation}
where $X_t \equiv A_t - \mu \cot \beta$, $\Delta_L \equiv (
\frac{1}{2} - \frac{2}{3} \, \sin^2 \theta_W ) m_Z^2 \cos 2 \beta $,
and $\Delta_R \equiv \frac{2}{3} \sin^2 \theta_W m_Z^2 \cos 2 \beta$.
Diagonalizing this matrix gives the physical masses of the lighter
stop $\tilde{t}_1$ and heavier stop $\tilde{t}_2$.  The radiative
contributions are maximized for heavy stops and large left-right
mixing with $X_t / M_S \approx \sqrt{6}$, where $M_S =
\sqrt{m_{\tilde{t}_1} m_{\tilde{t}_2}}$.  This ``maximal mixing''
relation is valid at 1-loop; it is modified by higher-order
corrections, but remains within $\sim 20\%$ of the 1-loop value.  For
$X_t \ll M_S$, however, conventional 2-loop analyses imply that the
measured Higgs mass requires stops with masses $\sim 5- 10~\tev$.  If
this is the characteristic mass scale of all squarks, they will be far
beyond the reach of the LHC or any near-future collider.

To improve the accuracy of current estimates of $m_h$, we use here the
program \hthreem~\cite{Kant:2010tf}.  Building on the 1- and 2-loop
terms provided by \feynhiggs~\cite{Frank:2006yh,Degrassi:2002fi,%
  Heinemeyer:1998np,Heinemeyer:1998yj}, \hthreem\ includes the roughly
16,000 diagrams that are the leading 3-loop corrections at
$\mathcal{O}(\alpha_t
\alpha_s^2)$~\cite{Kant:2010tf,Harlander:2008ju}.

When evaluating $m_h$, special care has to be taken to use accurate
numbers for the values of the input parameters entering the
calculation, most notably the top quark mass $m_t$ and the strong
coupling constant $\alpha_s$ in SUSY-QCD, renormalized in the \drbar{}
scheme (\ie, using dimensional reduction and modified minimal
subtraction), at a specific renormalization scale $\mu$.  These must
be calculated from the experimentally accessible values of the top
quark pole mass and $\alpha_s(m_Z)$ in five-flavor QCD.

In the original version of \hthreem, the transition of $m_t$ from the
on-shell to the \drbar{} scheme could suffer from large logarithms if
superpartners masses or renormalization scales $\mu$ are much larger
than $m_t$.  Since null results from the LHC increasingly favor this
possibility, the program has been improved in the following way.
First, we calculate $m_t(\mu)$ in five-flavor QCD in the \msbar{}
scheme using 4-loop running as implemented in the numerical package
RunDec~\cite{Chetyrkin:2000yt}.  This value is transferred to the
\drbar{} scheme via a finite renormalization at 3-loop
order~\cite{Harlander:2006xq,Jack:2007ni}.  Finally, the transition
from five-flavor QCD to SUSY-QCD is performed using the 2-loop
decoupling coefficient of $m_t$~\cite{Bauer:2008bj,mihaila:private}.
This procedure is faster, more robust, and more accurate than the old
code.  The new version of \hthreem\ is publicly available at
http://www.ttp.kit.edu/Progdata/ttp10/ttp10-23.

\ssection{Results as a Function of Weak-Scale Parameters.}  We now
present results for the Higgs boson mass, including the 3-loop
corrections described above, as functions of weak-scale supersymmetry
parameters.  We set $\tan\beta = 20$ so that the tree-level Higgs
boson mass is within 1 GeV of its maximal value, and we consider
nearly degenerate, unmixed stops, with $m_{\tilde{t}_L} =
m_{\tilde{t}_R}$ and $X_t = 0$.  The dependence on other parameters is
relatively mild; we set $\mu = 200~\gev$, assume gaugino mass
unification with $m_{\tilde{g}} = 1.5~\tev$, and set all other
sfermion soft mass parameters equal to $m_{\tilde{t}_{L,R}} + 1~\tev$.
For multi-TeV values of the sfermion masses, these models have scalar
masses far heavier than gaugino and Higgsino masses.

The results are shown in \figref{h3mresults}.  For $m_{\tilde{t}_1}$
in the range 1--10 TeV, 1-loop corrections raise the Higgs mass by 18
to 31 GeV, and 2-loop corrections raise the mass further by another 4
to 7 GeV.  The experimental value of $m_h$ is apparently obtained for
$m_{\tilde{t}_1} \sim 5~\tev$.  However, the 3-loop effects raise the
Higgs mass by another 0.5 to 3 GeV. The magnitude of the corrections
decreases with increasing loop order, indicating a well-behaved, if
slowly converging, perturbative expansion, and the size of the 3-loop
corrections is consistent, within uncertainties, with the NLL analysis
of Ref.~\cite{Martin:2007pg}.  Clearly, however, the 3-loop
corrections are still sizable, and they reduce the required top squark
mass to 3 to 4 TeV, a reduction with potentially great significance
for supersymmetry discovery, as we discuss below.

Ref.~\cite{Martin:2007pg} observes partial cancellations between
leading logarithm terms of $\mathcal{O}(\alpha_t \alpha_s^2)$ and
$\mathcal{O}(\alpha_t^2 \alpha_s)$ in a particular scenario.  We
advocate a full calculation at $\mathcal{O}(\alpha_t^2 \alpha_s)$ to
investigate whether this behaviour is universal.

\begin{figure}[tb]
\includegraphics[width=0.94\columnwidth]{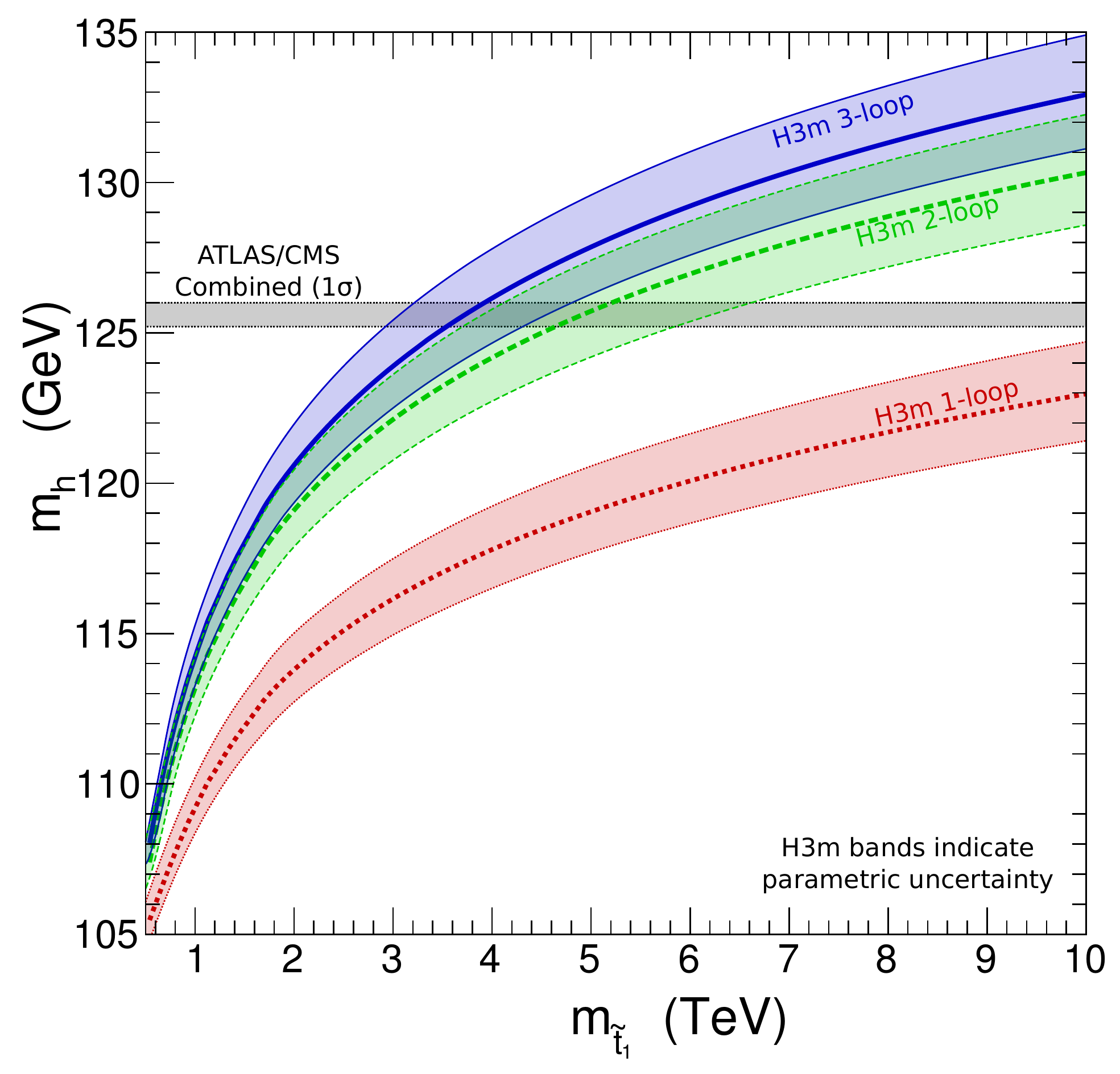}
\vspace*{-.1in}
\caption{The Higgs boson mass $m_h$ from \hthreem\ at 1-, 2-, and
  3-loops for nearly degenerate ($m_{\tilde{t}_L} = m_{\tilde{t}_R}$),
  unmixed ($X_t = 0$) top squarks, as a function of the physical mass
  $m_{\tilde{t}_1}$.  The renormalization scale is fixed to $M_S =
  \sqrt{m_{\tilde{t}_1} m_{\tilde{t}_2}}$, we set $\tan\beta = 20$,
  $\mu = 200~\gev$, all other sfermion soft parameters equal to
  $m_{\tilde{t}_{L,R}} + 1~\tev$, and assume gaugino mass unification
  with $m_{\tilde{g}} = 1.5~\tev$.  The bands indicate the parametric
  uncertainty from $m_t^{\text{pole}}=173.3 \pm 1.8~\gev$ and
  $\alpha_s(m_Z)=0.1184 \pm 0.0007$.  The horizontal bar is the
  experimentally allowed range $m_h = 125.6 \pm 0.4~\gev$.
\label{fig:h3mresults} 
}
\end{figure}

In \figref{h3mresults}, the width of the bands is determined by the
parametric uncertainty induced by the uncertainty in the top quark
mass and $\alpha_s$.  It is dominated by the uncertainty in the top
mass.  The top mass has been constrained by kinematic fits in combined
analyses of Tevatron~\cite{CDF:2013jga} and LHC~\cite{CMS:2012awa}
data, and may also be stringently constrained in the future by cross
section measurements (see, \eg, Ref.~\cite{Alioli:2013mxa}).  For now,
we consider the range $m_t^{\text{pole}}=173.3 \pm 1.8~\gev$.  The
resulting parametric uncertainty is 0.5 to 2 GeV; it exceeds the
experimental uncertainty and is comparable to that expected from 4-
and higher-loop effects in the theoretical prediction.

In \figref{comparison}, we compare our results to those of 2-loop
codes. The 2-loop results differ significantly from each other, with
differences of up to 4 GeV for stop masses in the 1 to 10 TeV range
shown.  The 3-loop results are within this range for $\sim \tev$ stop
masses, as found in Refs.~\cite{Kant:2010tf,Harlander:2008ju}.
However, for multi-TeV stop masses, the 3-loop contributions may
significantly enhance $m_h$.

\begin{figure}[tb]
\includegraphics[width=0.94\columnwidth]{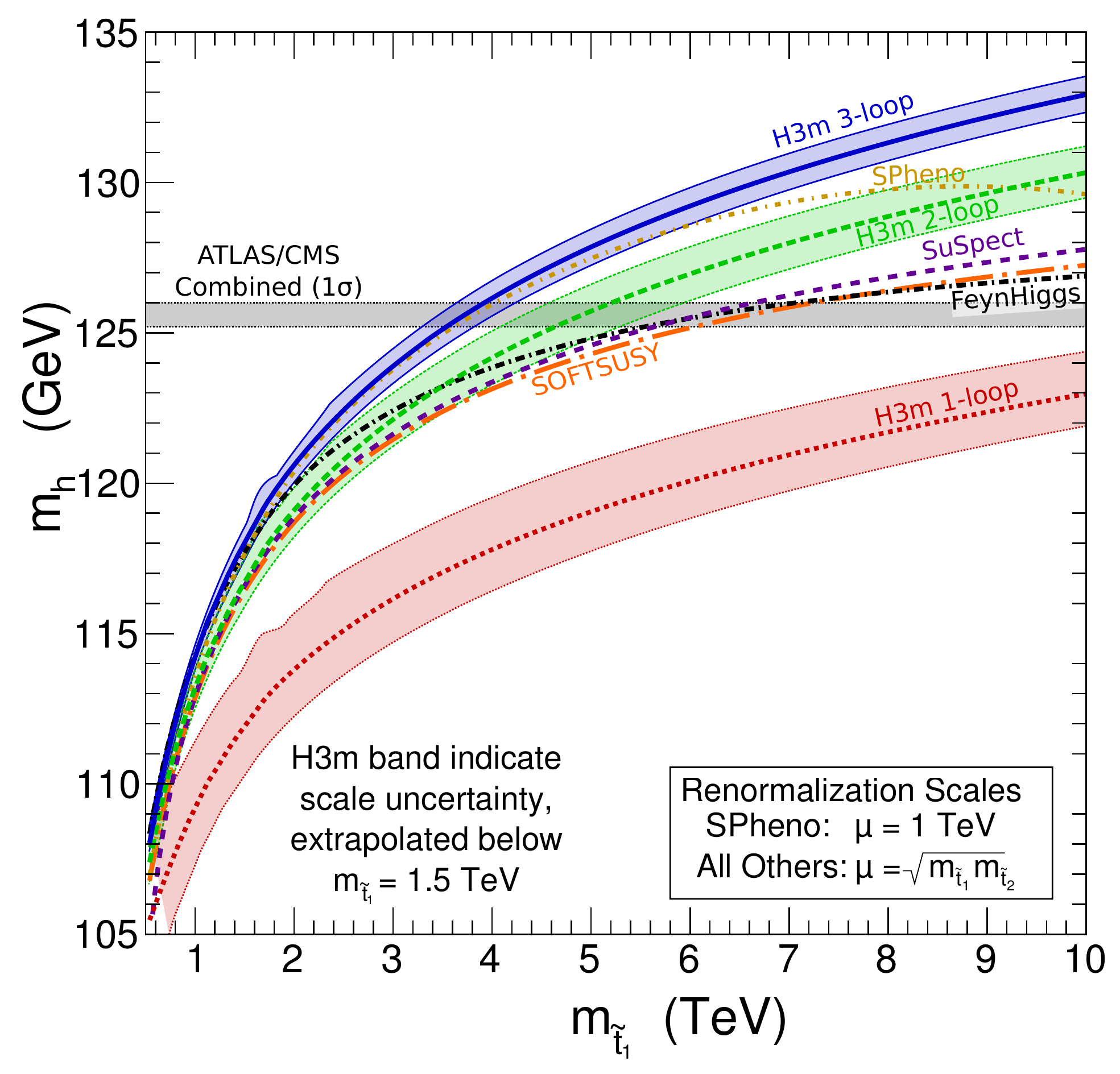}
\vspace*{-.1in}
\caption{ Comparison of \hthreem\ results with the 2-loop results of
  \feynhiggs~\cite{Heinemeyer:1998yj,Heinemeyer:1998np,%
    Degrassi:2002fi,Frank:2006yh}, \softsusy~\cite{Allanach:2001kg},
  \suspect~\cite{Djouadi:2002ze}, and
  \spheno~\cite{Porod:2003um,Porod:2011nf}.  The \hthreem\ bands
  indicate the uncertainty from varying the renormalization scale
  between $M_S/2$ and $2 M_S$.  The supersymmetry parameters are as in
  \figref{h3mresults}.
\label{fig:comparison} 
}
\end{figure}

Some of the differences between the 2-loop results can be explained by
different default choices for the renormalization scale.  They also
differ in how the running top mass is extracted from its pole mass.
This difference is formally of higher order~\cite{Allanach:2004rh}.
The different treatment of parameters also explains the difference
between \hthreem's 2-loop results and \feynhiggs.  For example,
\feynhiggs{} uses 1-loop running for $\alpha_s$ and $m_t$, which is
formally correct since the 2-loop results are leading order in
$\alpha_s$.

\ssection{Results for mSUGRA and Implications for Supersymmetry at the
  LHC.}  To determine the implications of the 3-loop corrections for
the LHC, we consider here the well-known framework of minimal
supergravity (mSUGRA), defined in terms of GUT-scale parameters, for
which detailed collider studies have been carried out.

In \figref{lhcreach} we show contours of $m_h$ with 3-loop corrections
in two well-studied $(m_0, M_{1/2})$ planes of mSUGRA.  To highlight
the regions of parameter space preferred by $m_h$, at each point in
parameter space, we define a theoretical uncertainty
$\Delta_{\text{th}} \equiv \sqrt{(\Delta_{\text{pert}})^2 
+ (\Delta_{\text{para}})^2}$, where
\begin{eqnarray}
\Delta_{\text{pert}} &\equiv& \frac{1}{2} \left| m_h^{\text{(3-loop)}} 
- m_h^{\text{(2-loop)}} \right| \;, \nonumber \\
\Delta_{\text{para}} &\equiv& \left| m_h (^{m_t =
  175.1~\gev}_{\alpha_s=0.1177}) 
- m_h(^{m_t = 173.3~\gev}_{\alpha_s=0.1184})
\right| . 
\end{eqnarray}
The quantity $\Delta_{\text{pert}}$ is the estimated uncertainty from
neglecting higher-order terms in the perturbation series.  It is
motivated by observing that the scale variation of the two-loop
prediction underestimates the 3-loop corrections, and is typically in
the 0.5 to 1.5 GeV range.  The parametric uncertainty
$\Delta_{\text{para}}$ arises dominantly from the uncertainty in the
top quark mass.  In the figure, we shade regions where the calculated
$m_h$ is within $\Delta_{\text{th}}$ and $2 \Delta_{\text{th}}$ of the
experimental central value 125.6 GeV.

\begin{figure}[tb!]
\includegraphics[width=0.94\columnwidth]{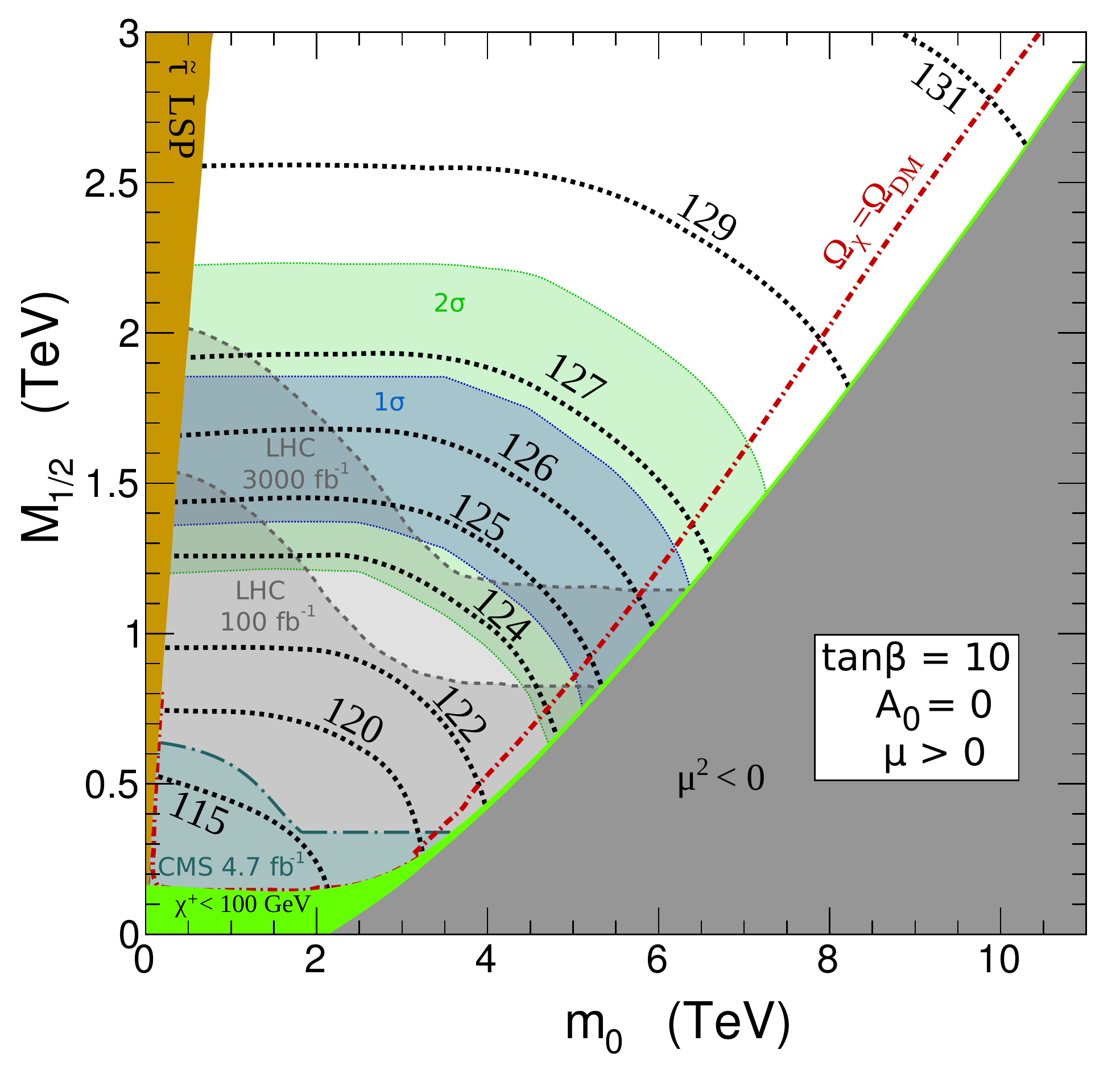}
\includegraphics[width=0.94\columnwidth]{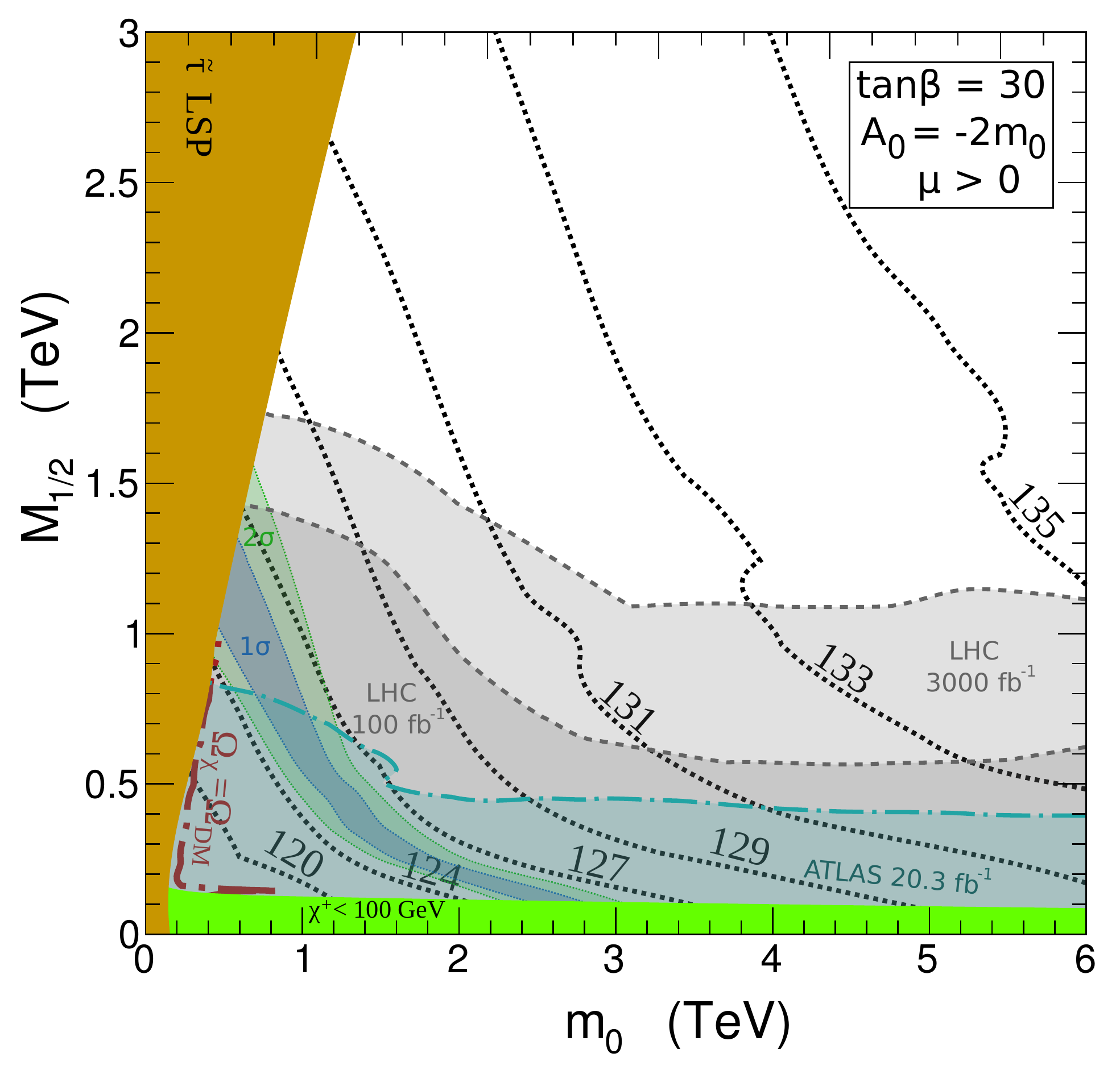}
\vspace*{-.1in}
\caption{3-loop \hthreem\ $m_h$ contours in two $(m_0, M_{1/2})$
  planes of mSUGRA, with $\tan\beta$, $A_0$, and $\text{sign}(\mu)$ as
  indicated.  In the dark blue (light green) shaded regions, the
  theoretical prediction is within $\Delta_{\text{th}}$
  (2$\Delta_{\text{th}}$) of the experimental central value.  On the
  $\Omega_{\chi} = \Omega_{\text{DM}}$ contour, thermal relic
  neutralinos are all the dark matter.  Top: Negligible stop mixing,
  with current exclusion contour from CMS~\cite{Chatrchyan:2012uea},
  and projected sensitivities of the 14 TeV LHC and its high-luminosity
  upgrade~\cite{Baer:2009dn}. Bottom: Significant stop mixing, with
  current exclusion contour from ATLAS~\cite{ATLAS-CONF-2013-047}, and
  projected sensitivities of the 14 TeV LHC and its high-luminosity
  upgrade~\cite{Baer:2012vr}.
\label{fig:lhcreach} 
}
\end{figure}

The positive 3-loop terms significantly impact the preferred range of
superpartner masses and the prospects for supersymmetry discovery at
the LHC.  In \figref{lhcreach}, top panel, $A_0 = 0$ and stop mixing
is negligible throughout the plane.  Requiring that the theoretical
prediction be within $2 \Delta_{\text{th}}$ of the experimental
central value, and imposing the further requirement that thermal relic
neutralinos make up all the dark matter (the focus point
region~\cite{Feng:1999mn,Feng:2000gh}), scalar mass parameters as low
as $m_0 \sim 4-5~\tev$, corresponding to stop masses as low as 3 to 4
TeV, and gluino masses as low as $m_{\tilde{g}} \simeq 2.8 M_{1/2}
\approx 2~\tev$ are consistent with the measured Higgs mass. These are
far lighter than the squark masses required if only 1- and 2-loop
corrections to $m_h$ are included.  Current bounds do not challenge
this parameter space~\cite{Chatrchyan:2012uea}, but the 14 TeV LHC
with $100~\ifb$ will already start probing the favored parameter
space, and a high-luminosity upgrade to $3~\text{ab}^{-1}$ may probe
most of it~\cite{Baer:2009dn}.  The LHC reach was extrapolated from a
study that used $\tan\beta = 45$~\cite{Baer:2009dn} by transferring
the $(m_{\tilde{q}}, m_{\tilde{g}})$ values on the reach contours to
the space with $\tan\beta = 10$.  The sensitivities are determined by
searches for multiple jets and missing energy along with a variable
number of leptons and are expected to be approximately independent of
$\tan\beta$.  Of course, lighter squark masses and brighter discovery
prospects are possible if one relaxes the cosmological requirement.

If there is significant stop mixing, the implications may be even more
dramatic.  This is illustrated in \figref{lhcreach}, bottom panel,
where $A_0 = -2 m_0$.  With the 3-loop corrections included, the
preferred region moves to $m_0$ as low as 1 TeV, and the $2\sigma$
region even overlaps the region with the correct thermal relic density
of neutralinos (the stau co-annihilation region).  Current
bounds~\cite{ATLAS-CONF-2013-047} exclude some of the favored region,
but the 14 TeV LHC with $100~\ifb$ will probe most of it, and it will
be explored fully by the LHC high-luminosity
upgrade~\cite{Baer:2012vr}.

\ssection{Conclusions.} 3-loop contributions to the Higgs boson mass
may be as large as 3 GeV in supersymmetric theories with multi-TeV
superpartners.  Given the extreme sensitivity of the stop mass to such
changes, this lowers the preferred stop mass to as low as 3 to 4 TeV,
with striking implications for supersymmetry discovery at the LHC.  In
models with a characteristic squark mass scale, these results imply
that even without significant mixing or additional particles, 1st and
2nd generation squarks may be within reach of the 14 TeV LHC with
$100~\ifb$, with much more promising prospects for a high-luminosity
upgrade.  Given the rapidly diminishing experimental uncertainty on
$m_h$, these results highlight the importance of improved theoretical
calculations of $m_h$, incorporating improved determinations of the
top quark mass, to refine the implications of the Higgs boson
discovery for supersymmetry.

\ssection{Acknowledgments.}  We thank S.~Heinemeyer, S.~P.~Martin,
L.~Mihaila, W.~Porod, P.~Slavich, M.~Steinhauser, X.~Tata, and N.~Zerf
for useful discussions, and P.~Draper for collaboration in early
stages of this work.  JLF is supported in part by U.S.~NSF grant
No.~PHY--0970173 and by the Simons Foundation.  PK is supported by the
DFG through SFB/TR-9 and by the Helmholtz Alliance ``Physics at the
Terascale.'' SP is supported in part by U.S.~DOE grant
No.~DE--FG02--04ER41268.  DS is supported in part by U.S.~DOE grant
No.~DE--FG02--92ER40701 and by the Gordon and Betty Moore Foundation
through Grant No.~776 to the Caltech Moore Center for Theoretical
Cosmology and Physics.

\end{document}